# Simple time-of-flight measurement of the speed of sound using smartphones

*S. Staacks, S. Hütz, H. Heinke, C. Stampfer*
*Institute of Physics I and II, RWTH Aachen University*

We propose an easy experiment that allows to determine the speed of sound through a simple time-of-flight measurement using two smartphones. The concept of using the sensors in mobile phones for physics experiments has become a well-known option for science teachers[1,2,3,4,5]. Since these devices are readily available to most students and teachers, experiments can be set up at little to no costs while generating fascination and motivation for the students thanks to the novelty of using their own devices in an unusual way. From all the sensors available in modern phones, the microphone offers the best temporal resolution by a huge margin as it records samples at a rate of (typically) 48 kHz. As this fast sampling rate is a requirement for high quality audio recordings it has been around long before smartphones offered other sensors, and audio inputs have already been used for experiments on soundcards in desktop computers[6,7,8].

Consequently, the microphone is at the center of a variety of phone-based experiments to determine the speed of sound. The experimental methods can generally be divided into time-of-flight measurements (often using echoes)[9,10] or resonance based measurements[11,12]. Unfortunately, both methods have their drawbacks. So far, time-of-flight measurement methods required students to extract the time from an app that is not designed to give this information directly, often by analyzing visual wave form representations of recorded audio signals, which can be difficult and too abstract for younger students. This quickly becomes a distraction from the core experiment and can be disproportionately time consuming. The resonance based experiments usually do not suffer from these problems as there are many apps which directly show the frequency spectrum of an audio recording and hence provide a numerical value of the peaks to work with. But these experiments require a higher level of understanding because the students need to know stationary waves and relate resonance frequencies to the geometry of the resonator in order to extract the speed of sound.

Our new experimental setup combines advantages of both methods by requiring only a minimum of physics background while still offering a simple numeric result directly from the app on the smartphone. This allows students to understand the principle of sound propagation and the speed of sound by calculating their result as the simple ratio of distance over time. For this experiment we use our free app "phyphox". To reach as many students as possible, we try to ensure that all of its features are available for the major operating systems if technically feasible for any device from Android 4.0 and iOS 8. The free app is available from the respective app stores. It is specifically designed for physics experiments in science education and offers access to the phone's sensors along with unique features such as customizable data analysis and a simple method to remotely access and control your experiment from a second device. For more information refer to http://phyphox.org.

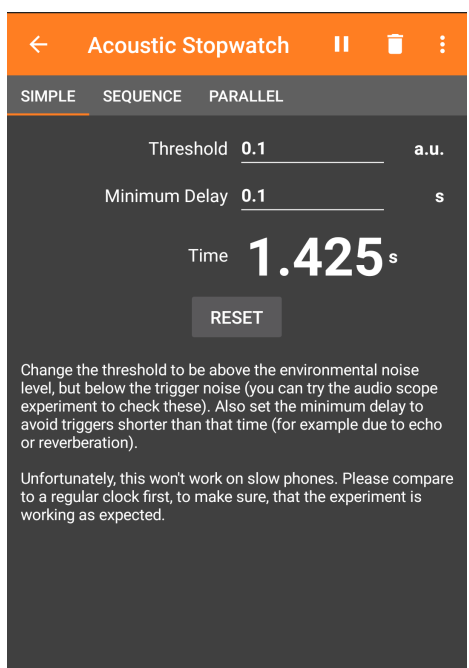

*Fig. 1: The app phyphox offers an acoustic stopwatch which measures the time between two acoustic events.*

Here, we use a specific data analysis tool of the app, called the "acoustic stopwatch" (Fig. 1). It uses the microphone to listen for sound to trigger the stopwatch. As soon as the audio amplitude exceeds a given threshold a time measurement is started. When the amplitude exceeds the threshold a second time, the measurement stops and the time between the two acoustic events is shown. Additionally, a minimum delay can be set to avoid unintentionally stopping the stopwatch with the first sound event due to its duration or reverberation. This concept can be shown to the students easily by clapping and letting them try out how the acoustic stopwatch reacts. If the stopwatch starts without clapping or misses claps, it is usually sufficient to increase the trigger threshold or to reduce it, which is the only experimental configuration required.

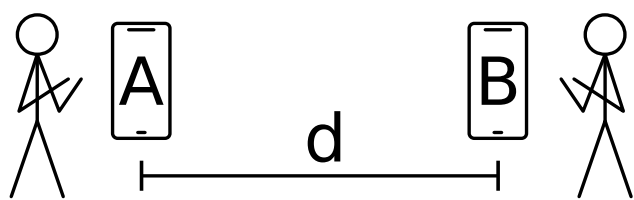

*Fig. 2: Student A generates an acoustic signal to start the timer on both phones. Student B then stops both with another acoustic signal. The phones measure different time intervals $\Delta t_A$ and $\Delta t_B$ granting access to the speed of sound.*

The experiment is conducted by two students with a smartphone each. They place the phones at a defined distance (for example 5 meters). Both students get their acoustic stopwatch ready and stand next to one of the phones each (Fig. 2). Then the first student (A) generates a loud acoustic signal (clapping, drums, popping balloons, etc.) to start the time measurement on both smartphones. Note, that the phone B at the other end will start its measurement at a delay equal to the time $\Delta t_d$ it takes for the sound to travel the distance d between both phones. Then, while both phones are running, the second student (B) generates a second acoustic signal to stop both phones. Again, the distant phone, which is in this case phone A, gets a delayed trigger. This means that phone A which started earlier by Δt_d, now also stops later by Δt_d. Therefore phone A measures a total time Δt_A equal to the time Δt_B measured by phone B plus two times Δt_d:

$$\Delta t_A = \Delta t_B + 2 \cdot \Delta t_d$$

Therefore, we can get Δt_d from the difference of both measured time intervals, resulting in the following equation for the speed of sound:

$$v = \frac{d}{\Delta t_d} = \frac{2d}{\Delta t_A - \Delta t_B}$$

The experimental results are surprisingly good. Since the microphone generates samples at a rate of 48 kHz, the theoretical temporal resolution is way below 1 ms. In practice, this is limited by finite risetimes of actual acoustical events which are recorded differently by both phones due to the attenuation over the distance. If you make sure that both phones and the origins of both trigger sounds are on a straight line and if you avoid wind, you can easily determine the speed of sound over a distance of 5 meters to an uncertainty of 30 m/s (357 m/s for our setup), which is quite accurate considering the simplicity of this experiment.

1. Vogt, Patrik, Jochen Kuhn, und Sebastian Müller. „Experiments Using Cell Phones in Physics Classroom Education: The Computer-Aided g Determination". The Physics Teacher 49, Nr. 6 (24. August 2011): 383–84. https://doi.org/10.1119/1.3628272.
2. Pendrill, Ann-Marie, und Johan Rohlén. „Acceleration and Rotation in a Pendulum Ride, Measured Using an IPhone 4". Physics Education 46, Nr. 6 (2011): 676. https://doi.org/10.1088/0031-9120/46/6/001.
3. Kuhn, Jochen, und Patrik Vogt. „Smartphones as Experimental Tools: Different Methods to Determine the Gravitational Acceleration in Classroom Physics by Using Everyday Devices". European Journal of Physics Education 4, Nr. 1 (2013): 16–27.
4. Chevrier, Joel, Laya Madani, Simon Ledenmat, und Ahmad Bsiesy. „Teaching classical mechanics using smartphones". The Physics Teacher 51, Nr. 6 (19. August 2013): 376–77. https://doi.org/10.1119/1.4818381.
5. Vieyra, Rebecca, Chrystian Vieyra, Philippe Jeanjacquot, Arturo Marti, und Martín Monteiro. „Turn Your Smartphone Into a Science Laboratory". The Science Teacher 82 (Januar 2015). https://doi.org/10.2505/4/tst15_082_09_32.
6. Mehrl, D., und M. Hagler. „Active Learning Using Inexpensive Sound Cards for Circuits and Communications Experiments". In Proceedings of the 28th Annual Frontiers in Education - Volume 03, 1102–1106. FIE '98. Washington, DC, USA: IEEE Computer Society, 1998. http://dl.acm.org/citation.cfm?id=1253527.1254152.
7. Carvalho, Carlos C., J. M. B. Lopes dos Santos, und M. B. Marques. „A Time-of-Flight Method To Measure the Speed of Sound Using a Stereo Sound Card". The Physics Teacher 46, Nr. 7 (5. September 2008): 428–31. https://doi.org/10.1119/1.2981293.
8. Aguiar, C. E., und M. M. Pereira. „Using the Sound Card as a Timer". The Physics Teacher 49, Nr. 1 (22. Dezember 2010): 33–35. https://doi.org/10.1119/1.3527753.
9. Kasper, Lutz, Patrik Vogt, und Christine Strohmeyer. „Stationary waves in tubes and the speed of sound". The Physics Teacher 53, Nr. 1 (29. Dezember 2014): 52–53. https://doi.org/10.1119/1.4904249.
10. Parolin, Sara Orsola, und Giovanni Pezzi. „Smartphone-aided measurements of the speed of sound in different gaseous mixtures". The Physics Teacher 51, Nr. 8 (8. Oktober 2013): 508–9. https://doi.org/10.1119/1.4824957.
11. Hirth, Michael, Jochen Kuhn, und Andreas Müller. „Measurement of sound velocity made easy using harmonic resonant frequencies with everyday mobile technology". The Physics Teacher 53, Nr. 2 (15. Januar 2015): 120–21. https://doi.org/10.1119/1.4905819.
12. Monteiro, Martín, Arturo C. Marti, Patrik Vogt, Lutz Kasper, und Dominik Quarthal. „Measuring the acoustic response of Helmholtz resonators". The Physics Teacher 53, Nr. 4 (23. März 2015): 247–49. https://doi.org/10.1119/1.4914572.